\documentclass{moriond}

\usepackage{xcolor}
\usepackage{amsmath,amssymb}
\usepackage{graphicx}

\def\be{\begin{equation}}
\def\ee{\end{equation}}
\def\bea{\begin{eqnarray}}
\def\eea{\end{eqnarray}}

\newcommand{\Photo}{}

\begin{document}
\vspace*{4cm}
\title{Looking for New Physics Through the Exclusive $b\to s\nu\bar{\nu}$ Modes}

\author{ Olcyr Sumensari }

\address{IJCLab, Pôle Théorie (Bat. 210), CNRS/IN2P3 et Université, Paris-Saclay, 91405 Orsay, France}

\maketitle

\abstracts{
The Belle-II experiment has recently measured $\mathcal{B}(B^+\to K^+\nu\bar{\nu})$, which appears to be almost $3\sigma$
 larger than its Standard Model (SM) prediction. In this talk, I will critically revisit the status of the SM predictions for the $B\to K^{(\ast)}\nu\bar{\nu}$ decays, and discuss the interpretation of the recent Belle-II measurement in terms of a general Effective Field Theory, as well as concrete models of physics beyond the SM.
 }

\section{Introduction}

Flavor Changing Neutral Currents (FCNCs) are powerful probes of New Physics (NP) effects since they are loop and CKM suppressed in the Standard Model (SM). The LHC experiments and the $B$-factories have precisely measured several observables related to the $b\to s\mu\mu$ transition in recent years~\cite{ParticleDataGroup:2022pth}. While some of these measurements display apparent deviations from the SM predictions~\cite{Alguero:2019ptt}, the interpretation of these results requires a precise understanding of hadronic uncertainties, in particular, those related to the so-called charm-loops that could hinder the sensitivity to NP effects~\cite{Ciuchini:2015qxb}.

Decays based on the $b\to s\nu\nu$ transition are very appealing, as they are theoretically cleaner than the ones with muons since they are unaffected by charm loops.  They are very sensitive to the effects of physics beyond the SM, thus offering a complementary ground to test, e.g.,~the NP scenarios proposed to explain the potential discrepancies in $b\to s\mu\mu$ decays. Furthermore, they also allow us to indirectly constrain models with large couplings to left-handed $\tau$'s due to $SU(2)_L$ gauge invariance, which are difficult to constrain otherwise~\cite{Allwicher:2023shc}. Recently, the Belle-II experiment observed, for the first time~\cite{Belle-II:2023esi},
\begin{equation}
    \mathcal{B}(B^+\to K^+\nu\bar{\nu})^\mathrm{exp} = [2.3 \pm 0.5(\mathrm{stat})^{+0.5}_{-0.4}(\mathrm{syst})]\times 10^{-5}\,,
\end{equation}
 which turns out to be $\approx 3\sigma$ above its SM prediction~\cite{Becirevic:2023aov}, namely $\mathcal{B}(B^+\to K^+\nu\bar{\nu})^\mathrm{SM}=(4.44\pm0.30) \times 10^{-6}$.~\footnote{The dependence of this result on the choice of CKM inputs will be discussed below. Note, also, that contributions from $B^+\to \tau^+ (\to K^+ \nu_\tau) \bar{\nu}_\tau$ are treated as a background in the experimental analysis and, for this reason, are removed from our SM predictions~\cite{Kamenik:2009kc}.} In the future, Belle II expects to improve this measurement, reaching an experimental precision of $\mathcal{O}(10\%)$ with $50~\mathrm{ab}^{-1}$ of data~\cite{Belle-II:2018jsg}. 
 
 In this talk, we will first critically revisit the current status of the SM predictions for the $B\to K^{(\ast)}\nu\bar{\nu}$ decays, emphasizing the theoretical improvements needed to match the future experimental sensitivity of Belle II~\cite{Becirevic:2023aov}.  Then, we will use this information to interpret the recent $B^+\to K^+\nu\bar{\nu}$ measurement in terms of a general Effective Field Theory (EFT) as well as concrete NP scenarios~\cite{Allwicher:2023xba}.

\section{EFT approach for $B\to K^{(\ast)}\nu\bar{\nu}$ decays}

Decays based on the $b\to s \nu \bar{\nu}$ transition are described by the following effective Lagrangian,
\begin{align}
\label{eq:eft-bsnunu}
\mathcal{L}_\mathrm{eff}^{\mathrm{b\to s\nu\nu}} =  \dfrac{4 G_F}{\sqrt{2}} \dfrac{e^2}{(4\pi)^2}\lambda_t \sum_{X=L,R} \sum_{i,j} C_X^{\nu_i\nu_j}\, (\bar{s} \gamma_\mu P_X b)(\bar{\nu}_i \gamma^\mu (1-\gamma_5)\nu_j)+\mathrm{h.c.}\,, 
\end{align}
\noindent where $G_F$ is the Fermi constant and $\lambda_t = V_{tb} V_{ts}^\ast$ is the product of the CKM-matrix entries. In the SM, only the left-handed Wilson coefficient appears,  $\big{[}C_{L}^{\nu_i\nu_j}\big{]}_\mathrm{SM}\equiv\delta_{ij}\, C_{L}^{\mathrm{SM}}$, which is precisely known~\cite{Buras:2014fpa},
\begin{align}
C_{L}^\mathrm{SM} = -X_t/\sin^2\theta_W\,, \qquad X_t=1.462(17)(2)\,,
\end{align}
\noindent where NLO QCD corrections have been included~\cite{Buchalla:1993bv}, 
as well as the two-loop electroweak contributions~\cite{Brod:2010hi}. Using $\sin^2\theta_W=0.23141(4)$ from PDG~\cite{ParticleDataGroup:2022pth}, one finally arrives at $C_{L}^{\mathrm{SM}}=-6.32(7)$, where the dominant source of uncertainty comes from the higher order QCD corrections. 

The effective Lagrangian~\eqref{eq:eft-bsnunu} can then be used to compute the $B\to K^{(\ast)}\nu\bar{\nu}$ branching fractions~\cite{Buras:2014fpa}, which will depend on two dominant sources of uncertainty: (i) the $B\to K^{(\ast)}$ form-factors and (ii) the CKM matrix elements, as we discuss below.

\subsection{Form-factors}

The $B\to K^{(\ast)}$ matrix elements are parameterized in full generality in terms of form factors which need to be computed non-perturbatively: 

 \paragraph{$\underline{B\to K}$:} The relevant hadronic matrix-element for the $B\to K$ transition reads
\begin{align}\label{eq:ff}
\langle \bar{K}(k) | \bar{s}\gamma^\mu b | \bar{B}(p) \rangle &= \Big{[}(p+k)^\mu- \dfrac{m_B^2-m_K^2}{q^2}q^\mu\Big{]} f_{+}(q^2)+ \dfrac{m_B^2-m_K^2}{q^2} q^\mu f_0(q^2)\,,
\end{align}
where $q^2=(p-k)^2$, and the vector ($f_+$) and scalar ($f_0$) form-factors satisfy the relation $f_+(0)=f_0(0)$. Only $f_+$ contributes to $B\to K\nu\bar{\nu}$ since neutrino masses are negligible. These form factors have been computed by means of Lattice QCD (LQCD) by the FNAL/MILC~\cite{Bailey:2015dka} and HPQCD collaborations~\cite{Parrott:2022rgu}. Following the same procedure as FLAG~\cite{Aoki:2021kgd}, we have updated the combined fit of these form factors~\cite{Becirevic:2023aov}, which supersedes the previous FLAG fit~\cite{Aoki:2021kgd}, in which now already obsolete HPQCD results have been used~\cite{Bouchard:2013eph}. By using these results, we can determine~\cite{Becirevic:2023aov}
\begin{align}
\label{eq:intB-SM-flag}
\mathcal{B}(B\to K & \nu\bar{\nu})^\mathrm{SM}/|\lambda_t|^2 =  
\left\{\begin{matrix}
(1.33 \pm 0.04)_{K_S}\times 10^{-3}\,, &  \\[0.5em]
(2.87 \pm 0.10)_{K^+}\times 10^{-3}\,, & 
\end{matrix}\right.
\end{align}

\noindent One should keep in mind, however, that these predictions result from an extrapolation since LQCD results are precisely obtained only for $q^2$ values, which is a potential source of systematic uncertainty. Therefore, it is fundamental to devise strategies to cross-check these results. A possibility is to measure the ratio of $\mathcal{B}(B\to K \nu\bar{\nu})$ at low- and high-$q^2$ bins~\cite{Becirevic:2023aov}, which would be a direct test of the form-factor $q^2$-shapes, since the CKM inputs, the $f_+$ form-factor normalization and also potential NP effects cancel out in such ratio.~\footnote{We assume neutrinos to be purely left-handed. }

 \paragraph{$\underline{B\to K^\ast}$:} The situation for the $B\to K^\ast$ transition is far more intricate than for $B\to K$ because there are more form factors in this case~\cite{Buras:2014fpa}. Furthermore, the results of only one LQCD study at nonzero recoil have been reported so far, with a specific lattice setup~\cite{Horgan:2013hoa}. 
In our study~\cite{Becirevic:2023aov}, we consider the combination~\cite{Bharucha:2015bzk} of the available LQCD values~\cite{Horgan:2013hoa} with those obtained by using the LCSR at low $q^2$, which then give
\begin{align}
\label{eq:intBst-SM}
\mathcal{B}(B\to K^{\ast } \nu\bar{\nu})^\mathrm{SM}/|\lambda_t|^2 = \left\{\begin{matrix}
(5.9 \pm 0.8)_{K^{\ast 0}}\times 10^{-3}\,, &  \\[0.5em]
(6.4 \pm 0.9)_{K^{\ast +}}\times 10^{-3}\,, & 
\end{matrix}\right.
\end{align}
We should point out, however, that this result is less robust than the one for $\mathcal{B}(B \to K  \nu\bar{\nu})$, due to our limited knowledge of the $B\to K^\ast$ form factors.

\subsection{CKM matrix elements}

The CKM factor $\lambda_t = V_{tb} V_{ts}^\ast$ introduces by far the largest parametric uncertainty in $\mathcal{B}(B\to K^{(\ast)} \nu\bar{\nu})$. The usual procedure, often adopted in the literature, is to determine $|V_{cb}|$ from tree-level processes and then to evaluate $|\lambda_t|$ using the CKM unitarity~\cite{Buras:2014fpa}.
Unfortunately, this procedure is intrinsically ambiguous since the CKM coupling $|V_{cb}^\mathrm{incl}|$ extracted from the inclusive semileptonic decay does not coincide with $|V_{cb}^\mathrm{excl}|$ obtained from the exclusive modes~\cite{Buras:2021nns}. The latest HFLAV~\cite{HFLAV:2022pwe} average values of the inclusive $|V_{cb}|$ are $|V_{cb}^\mathrm{incl}|_\mathrm{kin}=(42.2\pm 0.8)\times 10^{-3}$ or $|V_{cb}^\mathrm{incl}|_\mathrm{1S}=(42.0\pm 0.5)\times 10^{-3}$, which are larger than $|V_{cb}^{B\to D}|=(40.0\pm 1.0)\times 10^{-3}$ obtained after combining~\cite{Aoki:2021kgd} the experimental results on the exclusive $B\to D l\nu$ decays~\cite{HFLAV:2022pwe} (with $l=e,\mu$) with the LQCD form factors~\cite{MILC:2015uhg}. This discrepancy remains true if one compares the inclusive values with the one derived from $B\to D^\ast l\nu$ by HFLAV~\cite{HFLAV:2022pwe}, namely $|V_{cb}^{B\to D^\ast}|=(38.5\pm 0.7)\times 10^{-3}$.~\footnote{The $B\to D^\ast$ have been recently computed at nonzero recoil for the first time~\cite{FermilabLattice:2021cdg}. Note, also,
that a larger $|V_{cb}^{B\to D^\ast}|$ value has been recently advocated~\cite{Martinelli:2021myh}.} In other words, there is a discrepancy in $|\lambda_t|$ depending on the particular input considered~\cite{Becirevic:2023aov},
\begin{equation}
|\lambda_t| \times 10^3=
\left\{\begin{matrix}
41.4 \pm 0.8\,, &\quad\;\, {\small(B\to X_c l \bar{\nu})} \\[0.5em]
39.3 \pm 1.0\,, &\quad {\small(B\to D l \bar{\nu})} \\[0.5em]
37.8 \pm 0.7\,, &\quad\;\, {\small(B\to D^\ast l \bar{\nu})}
\end{matrix}\right.
\end{equation}

\noindent where the inclusive value is about $1\sigma$ and $2\sigma$ larger the ones derived from $B\to D l \bar{\nu}$ and $B\to D^\ast l \bar{\nu}$ decays, respectively, which again highlights the importance of clarifying the issue of the determination of $|V_{cb}|$.

\subsection{Summary}

By combining the form-factor and CKM determinations determined above, we obtain our final SM predictions at $\mathcal{O}(G_F^2)$, 
\begin{align}
    \mathcal{B}(B^+\to K^+ \nu \bar{\nu}) &= 4.44 (14)(27)\times 10^{-6}\,,  &\mathcal{B}(B^0\to K_S \nu \bar{\nu}) &=2.05(7)(12)\times 10^{-6}\,, \\[0.4em]
    \mathcal{B}(B^+\to K^{\ast +} \nu \bar{\nu}) &=9.8 (13)(6)\times 10^{-6}\,,  &  \mathcal{B}(B^0\to K^{\ast 0} \nu \bar{\nu}) &=9.1(13)(6)\times 10^{-6}\,,  \nonumber
\end{align}
\noindent where the first uncertainty comes from the form factors and the second one is dominated by the uncertainty of $|\lambda_t|$, where we have used $|\lambda_t|_\mathrm{excl}=(39.3\pm 1.0)\times 10^{-3}$ as a reference value~\cite{Becirevic:2023aov}. However, we stress again that by using $|\lambda_t|_\mathrm{excl}$ we obtain branching fractions that are about $1.5\sigma$ smaller than the values we would get by using $|\lambda_t|_\mathrm{incl}$. Therefore, it is fundamental to resolve this puzzle to match the future $\mathcal{O}(10\%)$ sensitivity of Belle II for these decays~\cite{Belle-II:2018jsg}.

\section{EFT description}

\subsection{Low-energy EFT}

The EFT contributions to $B\to K^{(\ast)}\nu\bar{\nu}$ can be expressed via the ratio
\begin{align}
\begin{split}
\dfrac{\mathcal{B}(B\to K^{(\ast)}\nu\bar{\nu})}{\mathcal{B}(B\to K^{(\ast)}\nu\bar{\nu})^\mathrm{SM}} =  \sum_{ij}\dfrac{|C_{L}^{\nu_i\nu_i}+ C_{R}^{\nu_i\nu_i}|^2}{3|C_{L}^\mathrm{SM}|^2} - \eta_{K^{(\ast)}}\sum_{i,j} \dfrac{\mathrm{Re}[C_R^{\nu_i\nu_j} C_L^{\nu_i\nu_j\,\ast}]}{3|C_{L}^\mathrm{SM}|^2}\,,
\end{split}
\end{align}
where $C_L^{\nu_i\nu_j} \equiv C_L^\mathrm{SM} \,\delta_{ij} +\delta C_L^{\nu_i\nu_j}$ and $C_R^{\nu_i\nu_j} \equiv \delta C_R^{\nu_i\nu_j}$. The last term in this expressions is absent for $B\to K \nu\bar{\nu}$ since $\eta_K=0$, but it contributes to $B\to K^\ast \nu\bar{\nu}$, where $\eta_{K^\ast} = 3.33(7)$ after integrating over the full $q^2$-range~\cite{Becirevic:2023aov,Allwicher:2023xba}. By using the above expressions, it is straightforward to derive the correlation between $\mathcal{B}(B\to K \nu\bar{\nu})$ and $\mathcal{B}(B\to K^\ast \nu\bar{\nu})$ which is depicted in the left panel of Fig.~\ref{fig:corr1}. While the left-handed scenario is strongly constrained by the upper limits on $\mathcal{B}(B\to K^\ast \nu\bar{\nu})$ from Belle data~\cite{Belle:2017oht}, the right-handed scenario can successfully explain the excess observed by Belle II, as it would deplete the $B\to K^\ast\nu\bar{\nu}$ rates~\cite{Becirevic:2023aov,Bause:2023mfe}. Another important prediction of the latter scenario is a sizable modification of the $B\to K^\ast\nu\bar{\nu}$ longitudinal polarization fraction, $F_L\equiv F_L(B\to K^\ast \nu\bar{\nu})$, which would be considerably lower than its SM value, $F_L^\mathrm{SM}=0.49(7)$, as depicted by $\mathcal{R}_{F_L} \equiv F_L/F_L^\mathrm{SM}$ in the right panel of Fig.~\ref{fig:corr1}~\cite{Becirevic:2023aov}. Therefore, these $B\to K^\ast \nu\bar{\nu}$ observables at Belle II would be useful tests of the observed deviation in $B\to K \nu\bar{\nu}$ decays.

\subsection{SMEFT}

We now assume that the NP scale lies well above the electroweak and write the EFT invariant under $SU(3)_c\times SU(2)_L \times U(1)$, namely the SMEFT~\cite{Buchmuller:1985jz}. Of all the $d=6$ operators in 
\begin{equation}
\mathcal L^{(6)}_\mathrm{SMEFT} \supset  \sum_i \frac{\mathcal{C}_i}{\Lambda^2} {\mathcal O}_i\,,
\end{equation}
we select those relevant to our study, namely, 
\begin{align}
\begin{split}\label{eq:4f}
\big{[}\mathcal{O}_{lq}^{(1)}\big{]}_{ijkl} &= \big{(}\overline{L}_i\gamma^\mu L_j\big{)} \big{(}\overline{Q}_k \gamma_\mu Q_l\big{)}\,, \qquad \big{[}\mathcal{O}_{lq}^{(3)}\big{]}_{ijkl} =\big{(}\overline{L}_i\gamma^\mu \tau^ I L_j\big{)} \big{(}\overline{Q}_k \tau^ I\gamma_\mu Q_l\big{)}\,,\\[0.4em]
\big{[}\mathcal{O}_{ld}\big{]}_{ijkl} &=\big{(}\overline{L}_i\gamma^\mu  L_j\big{)} \big{(}\overline{d}_{k R}\gamma_\mu d_{l R}\big{)}\,,
\end{split}
\end{align}
where $ Q$ and $L$ denote the quark and lepton $SU(2)_L$ doublet, respectively, while $u, d, e$ stand for the quark and lepton weak singlets, and $\lbrace i,j,k,l\rbrace$ denote flavor indices.~\footnote{We take the basis with a diagonal down-type quark Yukawa matrix, i.e.,~with  $Q_i=[(V^\dagger\,u)_{Li}\, ,\,d_{Li} ]^T$.} We will not consider the Higgs-current operators, since they are tightly constrained by $B_s-\overline{B_s}$ mixing and $\mathcal{B}(B_s\to \mu\mu)$~\cite{Allwicher:2023xba}. The tree-level matching between the SMEFT Lagrangian to Eq.~\eqref{eq:eft-bsnunu} gives 
\begin{align}
\begin{split}
\label{eq:left-CL-bsnunu1}
\delta C_L^{\nu_i \nu_j} &= \dfrac{\pi}{\alpha_\mathrm{em} \lambda_t} \dfrac{v^2}{\Lambda^2} \left\lbrace\big{[}\mathcal{C}_{lq}^{(1)}\big{]}_{ij23}-\big{[}\mathcal{C}_{lq}^{(3)}\big{]}_{ij23}\right\rbrace \,, \qquad 
\delta C_R^{\nu_i \nu_j} = \dfrac{\pi}{\alpha_\mathrm{em} \lambda_t} \dfrac{v^2}{\Lambda^2} \big{[}\mathcal{C}_{l d}\big{]}_{ij23}\,.
\end{split}
\end{align}

\begin{figure}[t!]
\centering
\centerline{\hspace{-5em}\includegraphics[width=0.595\linewidth]{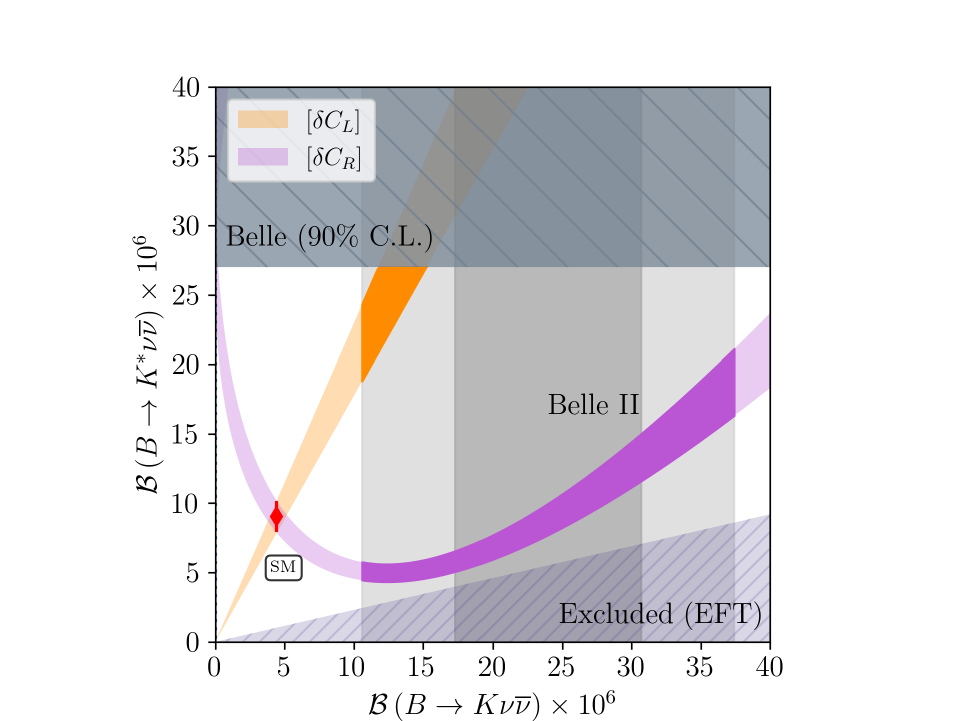}~\hspace{-3.6em}~\includegraphics[width=0.52\linewidth]{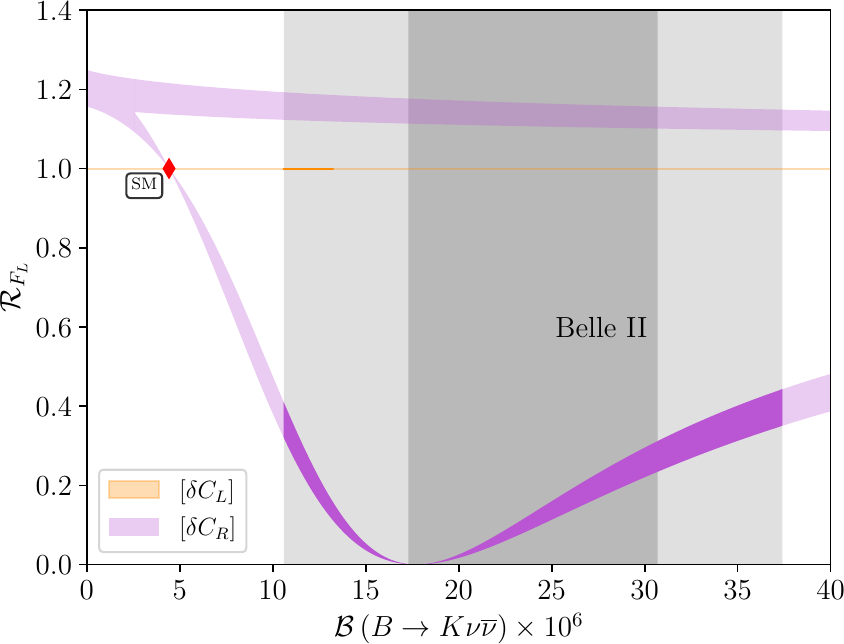}}
\caption{\small \sl The correlation between $\mathcal{B}(B\to K\nu\bar{\nu})$ with $\mathcal{B}(B\to K^{\ast}\nu\bar{\nu})$ (left panel) and $\mathcal{R}_{F_L} \equiv F_L/F_L^{\mathrm{SM}}$ (right panel) are shown with respect to the variation of $\delta C_{L}$ or $\delta C_{R}$.~\protect\cite{Becirevic:2023aov} The shaded gray area correspond to $1\sigma$ and $2\sigma$ of the recent Belle~II result for $\mathcal{B}(B\to K\nu\bar{\nu})$. The red point corresponds to the SM predictions for these observables. We also show the region of experimentally excluded $\mathcal{B}(B\to K^{\ast}\nu\bar{\nu})$ values~\protect\cite{Belle:2017oht} (gray hatched area), as well as the region not accessible within the EFT approach  (purple hatched area).}
\label{fig:corr1} 
\end{figure}

\noindent From Fig.~\ref{fig:corr1}, we find that the SMEFT coefficient that best describes current data is $\mathcal{C}_{ld}$. By considering a single lepton flavor of these coefficients ($i=j$), we find that Belle-II data is compatible with $|\mathcal{C}_{ld}|/\Lambda^2 \approx (5~\mathrm{TeV})^{-2}$. 

One of the advantages of considering the SMEFT is the correlation induced by $SU(2)_L$ invariance between $b\to s\nu_\ell\bar{\nu}_\ell$ and $b\to s\ell\ell$ modes, which allow us to probe the lepton flavor structure of these coefficients. Given the stringent constraints~\cite{ParticleDataGroup:2022pth} on $\mathcal{B}(B_s\to\mu\mu)$ and $R_{K^{(\ast)}}$, we concluded that a scenario that only affects electrons and muons cannot accommodate current data~\cite{Allwicher:2023xba}. Similarly, lepton flavor violating contributions ($i\neq j$) are tightly constrained by the upper limits on $\mathcal{B}(B_s\to \ell_i\ell_j)$ and $\mathcal{B}(B\to K^{(\ast)}\ell_i\ell_j)$. We find that the only scenario that can successfully accommodate experimental data has $\tau$-lepton coefficients~\cite{Allwicher:2023xba}. Notice that such EFT scenario is also compatible with LHC constraints on mono- and di-tau production at high-$p_T$~\cite{Allwicher:2022gkm}.

Finally, we comment on the realization of the above EFT scenario in terms of concrete models~\cite{Allwicher:2023xba,Bause:2023mfe}. The simplest possibility to generate $[\mathcal{C}_{ld}]_{3323}$ via the tree-level exchange of a heavy $Z^\prime \sim (\mathbf{1},\mathbf{1},0)$ or a scalar leptoquark $\widetilde{R}_2 \sim (\mathbf{3},\mathbf{2},1/6)$, where we denote the SM quantum numbers by $(SU(3)_c,SU(2)_L,U(1)_Y)$. However, the $Z^\prime$ scenario would be tightly constrained as it contributes to $\Delta m_{B_s}$ at tree level, in such a way that the couplings to $\tau$'s would have to be very large, being in potential tension with high-$p_T$ constraints. Instead, the $\widetilde{R}_2$ leptoquark can accommodate the large value of $\mathcal{B}(B\to K \nu\bar{\nu})$ while remaining compatible with other constraints, provided that its mass satisfies $m_{\widetilde{R}_2}\lesssim 3~\mathrm{TeV}$ to suppress its loop contribution to $B_s-\overline{B_s}$ mixing~\cite{Allwicher:2023xba}.

\subsection{Beyond the SMEFT}

Finally, we briefly comment on alternatives to explain the excess observed by Belle II, which exploit that $B\to K +\mathrm{inv}$ could also receive contributions from light invisible NP particles if they are sufficiently light to be produced on-shell. These particles could be a fermion, such as a right-handed neutrino~\cite{Felkl:2023ayn}, or light scalar/vector bosons~\cite{Altmannshofer:2023hkn} that could act as a portal to a dark sector. In both cases, our EFT approach has to be extended to account for these light degrees of freedom. The main prediction of these scenarios is a different $q^2$-distribution that would, e.g., be peaked at $q^2\simeq m_X^2$ for a two-body decay $B\to K X$, where $X$ is a light boson, which could be explored by a dedicated analysis at Belle II~\cite{Altmannshofer:2023hkn}.

\section{Summary and outlook}

The $B\to K^{(\ast)}\nu\bar{\nu}$ decays are powerful probes of NP effects as they are suppressed in the SM. The main interest in studying these decays is that they are theoretically cleaner than similar decays with muons, as they are not affected by the so-called charm loops. The main theoretical uncertainties of these decays are the hadronic factor and the parametric uncertainty stemming from the determination of the CKM input $\lambda_t=V_{tb} V_{ts}^\ast$. The latter is particularly problematic given the long-standing discrepancy in $|V_{cb}|$ from exclusive and inclusive decays, creating ambiguity in predicting $\lambda_t$ through CKM unitarity, which must be solved to match the expected sensitivity for these decays at Belle II.

Recently, the Belle-II experiment observed for the first time the $B^+\to K^+ \nu\bar{\nu}$ decays, with a branching fraction that is about $3\sigma$ larger than the SM predictions. We have shown that this deviation can be easily accommodated by an EFT with operators coupled to third-generation leptons, which is compatible with all available low- and high-energy constraints. Several measurements can be made to independently verify this excess, such as the measurement of $\mathcal{B}(B^0\to K_S \nu\bar{\nu})$, as well as the determination of the branching fraction and longitudinal-polarization asymmetry of $B\to K^\ast \nu\bar{\nu}$ decays, which can all be studied at Belle II. Finally, another possibility is that the excess is due to $B\to K X$, where $X$ denotes light and invisible particles. The experimental signature of such a scenario would be a different $q^2$-distribution kinematical distribution with respect to the SM, which could also be probed at Belle II.

\section*{Acknowledgments}

This project has received funding /support from the European Union’s Horizon 2020 research and innovation programme under the Marie Skłodowska-Curie grant agreement No 860881-HIDDeN and No 101086085-ASYMMETRY.

\end{document}